\documentclass[conference]{IEEEtran}
\IEEEoverridecommandlockouts

\usepackage{xcolor}
\usepackage{amsmath,amsfonts,amsthm}
\usepackage{algorithmic}
\usepackage{algorithm}
\usepackage{array}
\usepackage[caption=false,font=normalsize,labelfont=sf,textfont=sf]{subfig}
\usepackage{textcomp}
\usepackage{stfloats}
\usepackage{url}
\usepackage{verbatim,enumerate}
\usepackage{graphicx}
\usepackage{cite}
\usepackage{multirow} 
\newtheorem{theorem}{\bf Theorem}

\newtheorem{lemma}{\bf Lemma}

\usepackage[
top=0.75in,
bottom=1.1in,
left=1.0in,
right=1.0in,
]{geometry}

\def\BibTeX{{\rm B\kern-.05em{\sc i\kern-.025em b}\kern-.08em
    T\kern-.1667em\lower.7ex\hbox{E}\kern-.125emX}}

\begin{document}

\title{A Novel Formula for Solving Quadratic Equations over Binary Extension Fields
}

\author{\IEEEauthorblockN{1\textsuperscript{st} Leilei Yu}
\IEEEauthorblockA{\textit{Shenzhen Institute for Advanced Study} \\
\textit{University of Electronic Science \& Technology of China}\\
Shenzhen, China \\
yuleilei@uestc.edu.cn}
\and
\IEEEauthorblockN{2\textsuperscript{nd} Yunghsiang S. Han}
\IEEEauthorblockA{\textit{Shenzhen Institute for Advanced Study} \\
\textit{University of Electronic Science \& Technology of China}\\
Shenzhen, China \\
yunghsiangh@gmail.com}
\and
\IEEEauthorblockN{3\textsuperscript{rd} Pingping Li}
\IEEEauthorblockA{\textit{Shenzhen Institute for Advanced Study} \\
\textit{University of Electronic Science \& Technology of China}\\
Shenzhen, China \\
chinabai\_li@163.com}
\and
\IEEEauthorblockN{4\textsuperscript{th} Jiasheng Yuan}
\IEEEauthorblockA{\textit{Shenzhen Institute for Advanced Study} \\
\textit{University of Electronic Science \& Technology of China}\\
Shenzhen, China \\
yuanjsh@std.uestc.edu.cn}
}

\maketitle

\begin{abstract}
Solving quadratic equations over finite fields is a fundamental task in algebraic coding theory and serves as a key subroutine for computing the roots of cubic and quartic polynomials. 
Notably, any quadratic polynomial over binary extension fields can be transformed into the reduced form $x^2+x+c\in \mathbb{F}_{2^m}[x]$, for which existing formula-based methods rely on heavy exponentiation or case distinctions on $m$ (odd/even or powers of two), limiting uniformity and efficiency. This paper presents a unified, formula-based solution for all positive integers $m$ that uses only exclusive-OR operations (XORs). The approach leverages a Reed-Muller matrix characterization of evaluations and transforms the problem into computing a binary matrix-vector multiplication. The total cost is at most $m^2-2m+1$ XORs, and under parallelism, the latency is $\lceil \log_2 m\rceil$ XORs, making the method attractive for low-power, low-latency applications.
\end{abstract}


\section{Introduction}
{Finding} the roots of polynomials over finite fields is a fundamental task in algebraic coding theory. In Bose-Chaudhuri-Hocquenghem (BCH) and Reed-Solomon (RS) codes~\cite{lin2001error}, for example, the roots of the error-locator polynomial directly determine error positions. A widely used approach is the Chien search~\cite{chien1964}, which evaluates the polynomial at every field element and identifies points at which the evaluation vanishes. Although its hardware-friendly structure enables low latency and broad adoption, exhaustive evaluation incurs substantial computational redundancy and, consequently, avoidable power consumption.

It is classical that polynomials of degree at most four over the complex field admit formula-based solutions in terms of their coefficients. Analogously, certain low-degree polynomials over finite fields also permit direct, non-exhaustive root computation, with quadratic equations playing a central role. Particularly, roots of cubic and quartic polynomials can be reduced to solving associated quadratic instances (see, e.g., \cite{chen1982formulas}). Consequently, the unified, formula-based solution for quadratic polynomials developed here can be leveraged to obtain roots of cubic and quartic polynomials over $\mathbb{F}_{2^m}$. For brevity, this paper focuses on the quadratic case and omits higher-degree derivations.

Specifically, any quadratic polynomial $ay^2+by+d$ with $b\neq 0$ can be reduced to the canonical form $x^2+x+ad/b^2$ via the substitution $x=ay/b$, where the case $b=0$ is trivial. Hence, without loss of generality, this paper studies the reduced quadratic polynomial over the binary extension field $\mathbb{F}_{2^m}$,
\begin{equation}
	f(x)=x^2+x+c \in \mathbb{F}_{2^m}[x]\;.
\end{equation}
If $f(x)$ has roots in $\mathbb{F}_{2^m}$, then it must be two distinct roots, i.e., $x_0,x_1$, due to $x_0+x_1=1$.
It should be noted that the formula solution of $f(x)\in\mathbb{F}_{2^m}[x]$ was first proposed in~\cite{chen1982formulas}. 
However, the proposed method in~\cite{chen1982formulas} is not unified, as $m$ must be split into odd and even cases for separate treatment. Moreover, the even case needs to be further partitioned by $m\bmod 4$, complicating implementation. A different approach was later proposed in \cite{cherly1998solving}, which depends on finding an element whose trace function value is one.
Furthermore, \cite{walker1999new} proposed a formula to solve the reduced polynomial, but the proposed one applies only when $m$ is a power of two. 
Notably, all of the above works rely heavily on exponentiation in $\mathbb{F}_{2^m}$, in addition to field additions, which hampers efficiency and uniform implementation across all $m$.
As a comparison, this paper proposes a unified, formula-based solution for quadratic equations over $\mathbb{F}_{2^m}$, where $m$ is an arbitrary positive integer. Importantly, the proposed method requires only a small number of XOR operations, avoiding exponentiation.
The comparison details between the proposed method and previous methods are shown in TABLE~\ref{tab:1}.

The remainder of the paper is organized as follows: Sec.~\ref{sec:2} introduces the method and gives an example.
Sec.~\ref{sec:3} analyzes the computational complexity of the proposed method, and compares it with other methods. 
Finally, Sec.~\ref{sec:con} concludes this paper.

\section{Solving Quadratic Equations}\label{sec:2}

Throughout this paper, $\mathbb{N}$ denotes the set of $\{0,1,2,3,...\}$, and $\mathbf{0}$ denotes a size-adaptive zero vector or matrix.
Consider the binary extension field $\mathbb{F}_{2^m}$ for some $m>0$, and the basis of $\mathbb{F}_{2^m}$ is denoted by $(1, \alpha,..., \alpha^{m-1})$, where $\alpha$ is a primitive element of the field.
Each element $w_i, 0\leq i < 2^m,$ in $\mathbb{F}_{2^m}$ can be represented as 
\begin{equation}
	w_i=\sum_{j=0}^{m-1}i_j \cdot  \alpha^j, ~\text{where}~i=\sum_{j=0}^{m-1} i_j\cdot 2^j, i_j\in \{0,1\} \;.
\end{equation}
Note that the binary representation of each element $w_i, 0\leq i <2^m,$ can be succinctly represented as the binary vector $(i_0, i_1, ..., i_{m-1})\in \mathbb{F}_2^m $.
The following lemma is useful for solving the equations involved in this paper.
\begin{lemma}[\cite{yu2023reed}]\label{lem:1}
	For any $\ell\in \mathbb{N}$, if $\ell$ is  a power of two, then
	\begin{equation}
		(w_0^\ell\	w_1^\ell\ w_2^\ell	\ \cdots\ w_{2^m-1}^\ell)
		=\sum_{0\leq j<m} \alpha^{j\ell} \cdot R_m(2^j) ,\label{eq:1}
	\end{equation}
	where $R_m(2^j)$ denotes the $2^j$-th row (starting from zero-th row) of $R_m$ with $R_m$ being the Reed-Muller~(RM) matrix defined by 
	\begin{equation}
		R_{j+1}=\begin{pmatrix}
			R_{j}&R_j\\
			\mathbf{0}&R_j
		\end{pmatrix}, \forall j\in \mathbb{N},~\text{with}~R_0=(1) \;.
	\end{equation}
	
	\begin{proof}
		Since $\ell$ is a power of two, one can know from formulas (13) and (15) in \cite{yu2023reed} that $\begin{pmatrix}
			w_0^\ell&	w_1^\ell	&w_2^\ell	&\cdots&w_{2^m-1}^\ell
		\end{pmatrix}=E_{\ell}\cdot R_m$,
		where $E_{\ell}$ is the row vector whose $2^j$-th element is $\alpha^{j\ell}$ and the other elements are zero.
		The above formula gives \eqref{eq:1}.
		This completes the proof.
	\end{proof}
\end{lemma}

\subsection{Proposed Method}
Now consider the reduced quadratic polynomial over $\mathbb{F}_{2^m}$
\begin{equation}\label{eq:04}
	f(x)=x^2+x+c  \in \mathbb{F}_{2^m}[x] \;,
\end{equation}
which, if solvable over $\mathbb{F}_{2^m}$, has two distinct roots $x_0$ and $x_1$ satisfying $x_0+x_1=1$. One can obtain the following lemma.

\begin{lemma}\label{lem:2}
	If $(i_0, i_1, ..., i_{m-1})$ is the binary vector of the finite field element $w_i\in \mathbb{F}_{2^m}$, then $f(w_i)=0$ if and only if the following identity holds
{\small	\begin{equation}\label{eq:6}
		\begin{pmatrix}
			b_0(c)\\
			b_1(c)\\
			\vdots\\
			b_{m-1}(c)
		\end{pmatrix}=
		B \cdot 
		\begin{pmatrix}
			i_0\\
			i_1\\
			\vdots\\
			i_{m-1}
		\end{pmatrix}\;,
	\end{equation}}
	where $b_\ell(c), \forall c\in \mathbb{F}_{2^m},$ denotes the $\ell$-th bit of the binary vector of $c$, and $B$ denotes the $m\times m$ square matrix $(a_{\ell,j})_{0\leq \ell <m}^{{0\leq j <m}}$ with $a_{\ell,j}=b_{\ell}(\alpha^j + \alpha^{2j})$.
	
	\begin{proof}	
		To begin with, one can check the evaluations of $f(x)$ at all points over $\mathbb{F}_{2^m}$, i.e.,
		\begin{equation}
			\begin{aligned}
				&\begin{pmatrix}
					f(w_0)& f(w_1)& \cdots &f(w_{2^m-1})
				\end{pmatrix}\\
				=&
				\begin{pmatrix}
					c&  1& 1
				\end{pmatrix}
				\cdot
				\begin{pmatrix}
					1&	1	& 1&\cdots	&1\\
					w_0 &	w_1 	&w_2 	&\cdots&w_{2^m-1} \\
					w_0^2&	w_1^2	&w_2^2	&\cdots&w_{2^m-1}^2
				\end{pmatrix}\;.
			\end{aligned}
		\end{equation}
		By using Lemma~\ref{lem:1}, the following formula can be derived
{	\small	\begin{equation}\label{eq:5}
			\begin{aligned}
				&\begin{pmatrix}
					f(w_0)& f(w_1)& \cdots &f(w_{2^m-1})
				\end{pmatrix}\\
				=&
				\begin{pmatrix}
					c&  1& 1
				\end{pmatrix}
				\cdot
				\begin{pmatrix}
					R_{m}(0)\\
					\sum_{0\leq j<m} \alpha^j \cdot R_m(2^j) \\
					\sum_{0\leq j<m} \alpha^{2j} \cdot R_m(2^j)
				\end{pmatrix} \\
				= &c\cdot R_{m}(0)+ \sum_{0\leq j<m}(\alpha^j+ \alpha^{2j} )\cdot R_m(2^j) 
			\end{aligned}\;,
		\end{equation}}
		where $R_m$ and $R_m(2^j)$ are defined in Lemma~\ref{lem:1}, and $R_m(0)$ denotes the first row of $R_m$.
		The above formula is similar to the encoding formula of first-order Reed-Muller (RM) codes~(please refer to \cite{dumer2006soft,reeves2023reed} for details).
		
		In order to align with the first-order RM codes, one can map \eqref{eq:5} into the case over the binary field. Then, \eqref{eq:5} can be rewritten as the following binary matrix form,
{	\small	\begin{equation}\label{eq:7}
			\begin{pmatrix}
				b_0(c)\cdot R_{m}(0)+ \sum_{j=0}^{m-1}b_0(\alpha^j+ \alpha^{2j} )\cdot R_m(2^j) \\
				b_1(c)\cdot R_{m}(0)+  \sum_{j=0}^{m-1}b_1(\alpha^j+ \alpha^{2j} )\cdot R_m(2^j) \\
				\vdots\\
				b_{m-1}(c)\cdot R_{m}(0)+ \sum_{j=0}^{m-1}b_{m-1}(\alpha^j+ \alpha^{2j} )\cdot R_m(2^j) 
			\end{pmatrix}\;,
		\end{equation}}
		where the $\ell$-th row, $ \forall 0\leq \ell < m$, is exactly the codeword of the first-order RM code corresponding to the message vector 
		\begin{equation}\label{eq:8}
			(b_\ell(c), b_\ell(1+ 1), b_\ell(\alpha+ \alpha^{2} ),  ...,b_\ell(\alpha^{m-1}+ \alpha^{2(m-1)} )) \;.
		\end{equation}
		From \cite{dumer2006soft,reeves2023reed}, the message polynomial corresponding to \eqref{eq:8} is as follows: (each $i_j, 0\leq j<m,$ represents an independent variable)
		\begin{equation}\label{eq:4}
			\begin{aligned}
				&g_{\ell}(i_0, i_1, ...,i_{m-1}) \\
				= &b_\ell(c) +  \sum_{0\leq j <m} b_\ell (\alpha^j+ \alpha^{2j} ) \cdot i_j, ~ \text{where}~i_j\in\{0,1\} \;.
			\end{aligned}
		\end{equation}
		Due to the fact that the RM codeword can be regarded as evaluating the message polynomial at all possible points,
		then \eqref{eq:7} can be rewritten as
{	\small	\begin{equation}\label{eq:10}
			\begin{pmatrix}
				g_{0}(0) & g_0(1) & g_{0}(2)&\cdots &g_0(2^m-1)\\
				g_{1}(0) & g_1(1) & g_{1}(2)&\cdots &g_1(2^m-1)\\
				\vdots\\
				g_{m-1}(0) & g_{m-1}(1) & g_{m-1}(2)&\cdots &g_{m-1}(2^m-1)
			\end{pmatrix}\;,
		\end{equation}}
		where the input $m$-tuple $(i_0, i_1, ...,i_{m-1})$ of $g(\cdot)$ is denoted by $i=\sum_{j=0}^{m-1}i_j\cdot 2^j$ for simplicity.

		Now, if $f(w_i)=0$ for some $0\leq i <2^m$, then the $i$-th column of the matrix in \eqref{eq:10} is a zero vector, which results in the identity of $g_\ell(i)=0, \ell=0,1,...,m-1$.
		According to \eqref{eq:4}, the $m$ identities leads to
{	\small	\begin{equation}\label{eq:15}
			\begin{aligned}
				&	\begin{pmatrix}
					b_0(c)\\
					b_1(c)\\
					\vdots\\
					b_{m-1}(c)
				\end{pmatrix}+B \cdot 
				\begin{pmatrix}
					i_0\\
					i_1\\
					\vdots\\
					i_{m-1}
				\end{pmatrix}=\mathbf{0}  ,
			\end{aligned}
		\end{equation}}
		where $B  =(b_{\ell}(\alpha^j + \alpha^{2j}))_{0\leq \ell <m}^{{0\leq j <m}}$ is the matrix of size $m\times m$. Thus, the above has completed the proof of necessity.

		To prove the sufficiency, assuming \eqref{eq:15} holds for some $(i_0,i_1,...,i_{m-1})$, then $g_\ell(i_0,i_1,...,i_{m-1})=0,\forall \ell\in [m],$ must hold.
		Based on the relationship between RM codewords and polynomial evaluation, the $i$-th element of the result in \eqref{eq:5} must be zero, where $i=\sum_{j=0}^{m-1}i_j\cdot 2^j$.
		This results in $f(w_i)=0$.
		This completes the proof.
	\end{proof}

\end{lemma}


Notably, the matrix $B$ in Lemma~\ref{lem:2} is independent of $f(x)$, so it can be pre-calculated, and one can record the process of transforming $B$ into its reduced row echelon form.
Let $P$ be the row transformation matrix that transforms $B$ into reduced row echelon form.
Since the first column of $B$ is a zero column and $B$ has the rank of $m-1$~(which is due to the equation of \eqref{eq:6} having at most two distinct roots), the reduced row echelon form of $B$ must be
\begin{equation}\label{eq:13}
	(\mathbf{0} | I_0)=P\cdot B \;,
\end{equation} 
where $I_0$ is the $m\times (m-1)$ matrix obtained by inserting one all-zero row to an $(m-1)\times (m-1)$ identity matrix.
Now, one can obtain a necessary and sufficient condition for the reduced quadratic equation to have a solution, as follows:
\begin{theorem}\label{thm:1}
	Given the matrix pair $(P, I_0)$ for $\mathbb{F}_{2^m}$, which is shown in \eqref{eq:13} and where the $\ell$-th row of $I_0$ is an all-zero row, $f(x)$ has roots in $\mathbb{F}_{2^m}$ if and only if the $\ell$-th element of the following result is zero:
{\small	\begin{equation}\label{eq:-14}
		\mathbf{s} =P \cdot 	\begin{pmatrix}
			b_0(c)\\
			b_1(c)\\
			\vdots\\
			b_{m-1}(c)
		\end{pmatrix}  \;,
	\end{equation}}
	where $\mathbf{s}$ is the vector of size $m\times 1$.
	\begin{proof}
		First, let the condition hold to prove its sufficiency.
		According to Lemma~\ref{lem:2}, it is sufficient to prove that
{	\small	\begin{equation}\label{eq:-15}
			\mathbf{s}=P\cdot B\cdot 		\begin{pmatrix} 
				i_0\\
				i_1\\
				\vdots\\
				i_{m-1}
			\end{pmatrix}=
			I_0\cdot 
			\begin{pmatrix} 
				i_1\\
				\vdots\\
				i_{m-1}
			\end{pmatrix} 
		\end{equation}}
		has a solution on $i_1, i_2,\ldots, i_{m-1}$. Since the $\ell$-th rows of $\mathbf{s}$ and $I_0$ are all zeros, 
		$(i_1, i_2,..., i_ {m-1}) $ can easily be obtained. 
		Thus, $f(x)$ must have roots. 
		Conversely, if $f(x)$ has roots and $f(w_i)=0$, then 
		\eqref{eq:6} in Lemma~\ref{lem:2} holds, which gives \eqref{eq:-15}.
		Since the $\ell$-th row of $I_0$ is an all-zero row, the $\ell$-th element of $\mathbf{s}$ must be zero.
		This completes the proof.
		
	\end{proof}
	
\end{theorem}

\begin{table*}[t]
	\centering
	\caption{Solving the reduced quadratic polynomial $f(x)=x^2+x+c \in \mathbb{F}_{2^m}[x]$ in $\mathbb{F}_{2^m}$\\(Binary representation of $c$ is $(b_0(c), b_1(c),...,b_{m-1}(c))$, the two roots are respectively $x_0=w_{\sum_{j=1}^{m-1}i_j\cdot 2^j}$ and $x_1=x_0+1$).}
	\label{tab:0}
{	\begin{tabular}{c|c||c}
		\hline
		$m$ & Primitive Polynomial & Solvability Criteria and Solution Formula, i.e., $P\cdot (b_0(c),..., b_{m-1}(c))^{\mathrm{T}}=I_0\cdot (i_1,...,i_{m-1})^{\mathrm{T}}$\\ \hline  \hline  
		3 & $x^3+x+1$ &   $	
		\begin{pmatrix}
			1&&\\
			&&1\\
			&1&1
		\end{pmatrix}\cdot \begin{pmatrix}
			b_0(c)\\
			b_1(c)\\
			b_2(c)
		\end{pmatrix}=
		\begin{pmatrix}
			&   \\
			1&    \\
			& 1
		\end{pmatrix}\cdot 
		\begin{pmatrix}
			i_1\\
			i_2
		\end{pmatrix}$  \\   \hline
		4 & $x^4+x+1$ &  $ 
		\begin{pmatrix}
			&&&1\\
			1&1&&\\
			1&&&\\
			&1&1&
		\end{pmatrix}\cdot 
		\begin{pmatrix}
			b_0(c)\\
			b_1(c)\\
			b_2(c)\\
			b_3(c)
		\end{pmatrix}=
		\begin{pmatrix}
			&&\\
			1 &  &\\
			& 1 &  \\
			& &1\\
		\end{pmatrix}\cdot 
		\begin{pmatrix}
			i_1\\
			i_2\\
			i_3
		\end{pmatrix}$  \\   \hline
		5 & $x^5+x^2+1$ &  
		$
		\begin{pmatrix}
			1&&&1&\\
			&&1&&1\\
			&&&1&1\\
			&1&1&&1\\
			&&&1&
		\end{pmatrix}\cdot 
		\begin{pmatrix}
			b_0(c)\\
			b_1(c)\\
			b_2(c)\\
			b_3(c)\\
			b_4(c)
		\end{pmatrix}=
		\begin{pmatrix}
			&&&\\
			1 & &&\\
			& 1 & & \\
			& &1&\\
			&&&1
		\end{pmatrix}\cdot 
		\begin{pmatrix}
			i_1\\
			i_2\\
			i_3\\
			i_4
		\end{pmatrix}$   \\   \hline  
		6 & $x^6+x+1$ &  $
		\begin{pmatrix}
			& & & & &1\\
			1 & 1 &   & & & \\
			& 1& 1 &1 &  & \\
			1 &   &  & & & \\
			1& & &1& & \\
			1&1&1& &1& 
		\end{pmatrix}\cdot 
		\begin{pmatrix}
			b_0(c)\\
			b_1(c)\\
			b_2(c)\\
			b_3(c)\\
			b_4(c)\\
			b_5(c)
		\end{pmatrix}=
		\begin{pmatrix}
			& & & & \\
			1 &   & & & \\
			& 1 &  &  & \\
			&  &1& & \\
			& & &1& \\
			& & & &1
		\end{pmatrix}\cdot 
		\begin{pmatrix}
			i_1\\
			i_2\\
			i_3\\
			i_4\\
			i_5
		\end{pmatrix}$     \\    \hline
		7 & $x^7+x^3+1$ &  $
		\begin{pmatrix}
			1& & & & & & \\
			&   & 1&1&1& &1\\
			&  &   &1 &1 & &1\\
			&   &  & & &1&1\\
			&1&1& &1& & \\
			& & &1& &1&1\\
			& & &1& & &1
		\end{pmatrix}\cdot 
		\begin{pmatrix}
			b_0(c)\\
			b_1(c)\\
			b_2(c)\\
			b_3(c)\\
			b_4(c)\\
			b_5(c)\\
			b_6(c)
		\end{pmatrix}=		\begin{pmatrix}
			& & & & & \\
			1 &   & & & & \\
			& 1 &  &  & & \\
			&  &1& & & \\
			& & &1& & \\
			& & & &1& \\
			& & & & &1
		\end{pmatrix}\cdot 
		\begin{pmatrix}
			i_1\\
			i_2\\
			i_3\\
			i_4\\
			i_5\\
			i_6
		\end{pmatrix}$   \\   \hline 
		8 & $x^8+x^4+x^3+x^2+1$ &{   $
		\begin{pmatrix}
			& & & & &1& & \\
			1 &   & 1 & &1& & & \\
			1&  &   &1&1 & &1& \\
			& 1 & 1&1&1& & & \\
			1& & & & & & &1\\
			&1&1&1&1& &1& \\
			1&1&1& &1& & &1\\
			1&1&1& &1& & & 
		\end{pmatrix}\cdot 
		\begin{pmatrix}
			b_0(c)\\
			b_1(c)\\
			b_2(c)\\
			b_3(c)\\
			b_4(c)\\
			b_5(c)\\
			b_6(c)\\
			b_7(c) 
		\end{pmatrix}=	
		\begin{pmatrix}
			& & & & & & \\
			1 &   & & & & & \\
			& 1 &  &  & & & \\
			&  &1& & & & \\
			& & &1& & & \\
			& & & &1& & \\
			& & & & &1& \\
			& & & & & &1
		\end{pmatrix}\cdot 
		\begin{pmatrix}
			i_1\\
			i_2\\
			i_3\\
			i_4\\
			i_5\\
			i_6\\
			i_7
		\end{pmatrix}$  }  \\
		\hline      
		\hline                                         
	\end{tabular}}
\end{table*}

\begin{table*}[t]
	\centering
	\caption{Comparison of solving $f(x)=x^2+x+c=0$ over $\mathbb{F}_{2^m}$.}
	\label{tab:1}
	\begin{tabular}{c||c|c|l}
		\hline
		Methods & $m$ & Formula-based Solutions & \multicolumn{1}{c}{Number of Operations}\\ \hline  \hline  
		\cite{chien1964}  & arbitrary  & None (using exhaustive search). &Field Add, Mul: $2^{m+1}$, $2^m$  \\ \hline
		\multirow{4}{*}{\cite{chen1982formulas}} & \multirow{2}{*}{odd } & $x_0=\sum_{j\in J}c^{2^j}=\sum_{i\in I}c^{2^i}$, where & Field Exponentiation: $(m-1)/2$ \\
		&  &$J=\{0,2,4,...,m-1\}$ and $I=\{1,3,5,...,m-2\}$. & Field Add: $(m-3)/2$ \\ \cline{2-4}
		&  \multirow{2}{*}{even } &There are formulas corresponding to distinct  & Exponentiation: $> (m-2)/2$\\
		&  &$Tr_4(c)$, where $Tr_4(c):=\sum_{i=0}^{(m-2)/2}c^{2^{2i}}$. &Field Add: $> (m-2)/2$ \\
		\hline

		\multirow{2}{*}{\cite{cherly1998solving}}  &  \multirow{2}{*}{arbitrary}  &$x_0=\sum_{j=1}^{m-1}c^{2^j} \cdot \left( \sum_{\ell=0}^{j-1}u^{2^{\ell}} \right)$,where $u\in \mathbb{F}_{2^m}$ and  & Field Exponentiation: $m-1$ \\
		&   &$Tr_2(u)=1$ with $Tr_2(u):=\sum_{i=0}^{m-1}u^{2^i}$. & Field Add, Mul: $m-2$, $m-1$ \\  \hline    
		
		\multirow{3}{*}{\cite{walker1999new}}  &  \multirow{3}{*}{power of two}  &\multirow{3}{*}{\text{There are formulas corresponding to distinct $m$.}}
		& Field Mul: $2$   ($m=2$) \\ 
		&    &   &   Field Add, Mul: $2$, $5$ ($m=4$) \\
		&    &  &  Field Exponentiation, Add, Mul: $2$, $4$, $7$  ($m=8$) \\        \hline 
		\multirow{2}{*}{Proposed}  & \multirow{2}{*}{arbitrary}   &Multiply the matrix $P$ by the binary vector of $c$, & XOR: $m^2-2m+1$\\
		&   &  as shown in Theorem~\ref{the:1}. &   (equiv. to $\leq m-1$ field additions)  \\ 
		\hline                                         
	\end{tabular}
\end{table*}

\begin{theorem}\label{the:1} 
	Given the matrix pair $(P, I_0)$ for $\mathbb{F}_{2^m}$, which is shown in \eqref{eq:13}, then $f(w_i)=0$ and $f(w_i+1)=0$, where $i=\sum_{j=1}^{m-1}i_j\cdot 2^j$ satisfies 
{\small	\begin{equation}\label{eq:17}
		P \cdot 	\begin{pmatrix}
			b_0(c)\\
			b_1(c)\\
			\vdots\\
			b_{m-1}(c)
		\end{pmatrix} =	I_0\cdot 
		\begin{pmatrix} 
			i_1\\
			\vdots\\
			i_{m-1}
		\end{pmatrix}\;.
	\end{equation}}
	
	\begin{proof}
		This follows directly from Lemma~\ref{lem:2}, \eqref{eq:13}, and the fact that $i_0$ can be either zero or one.
		This completes the proof.
	\end{proof}
	
\end{theorem}


\subsection{Example}
Let $m=7$ and the primitive polynomial of $\mathbb{F}_{2^m}$ be $x^7+x^3+1$. Then $\alpha$ in $\alpha^7=\alpha^3+1$ is a primitive element.
In this case, the matrix $B$ in \eqref{eq:6} is
{\small \begin{equation}
	B=			\begin{pmatrix}
		0 & 0 & 0&0 & 0 & 0 &0\\
		0 & 1 & 0&0 & 1 & 0 &1\\
		0 & 1 & 1&0 & 0 & 0 &0\\
		0 & 0 & 0&1 & 0 & 1 &0\\
		0 & 0 & 1&0 & 0 & 0 &1\\
		0 & 0 & 0&0 & 0 & 1 &1\\
		0 & 0 & 0&1 & 0 & 1 &1
	\end{pmatrix}  \;.
\end{equation}}
It can be transformed into the reduced row echelon form	
{\small \begin{equation}
	\begin{pmatrix}
		0 & 0 & 0&0 & 0 & 0 &0\\
		0 & 1& 0&0 & 0 & 0 &0\\
		0 & 0 & 1&0 & 0 & 0 &0\\
		0 & 0 & 0&1 & 0 & 0 &0\\
		0 & 0 & 0&0 & 1 & 0 &0\\
		0 & 0 & 0&0 & 0 & 1 &0\\
		0 & 0 & 0&0 & 0 & 0 &1
	\end{pmatrix} =(\mathbf{0} | I_0)=P\cdot B 
\end{equation}
through some elementary row transformations.
Obviously, the matrix $I_0$ is the matrix obtained by adding an all-zero row on top of the $6\times 6$ identity matrix.
In addition, the above process leads to the row transformation matrix $P$ in the matrix pair $(P, I_0)$ being
{\small \begin{equation}
	P=			\begin{pmatrix}
		1 & 0 & 0&0 & 0 & 0 &0\\
		0 & 0& 1&1 & 1 & 0 &1\\
		0 & 0 & 0&1 & 1 & 0 &1\\
		0 & 0 & 0&0 & 0 & 1 &1\\
		0 & 1 & 1&0 & 1 & 0 &0\\
		0 & 0 & 0&1 & 0 & 1 &1\\
		0 & 0 & 0&1 & 0 & 0 &1
	\end{pmatrix}  \;.
\end{equation}}
According to the formula solution in Theorem~\ref{the:1}, 
\begin{equation}\label{eq:21}
	P \cdot 	\begin{pmatrix}
		b_0(c)\\
		b_1(c)\\
		\vdots\\
		b_{6}(c)
	\end{pmatrix}=
	\begin{pmatrix}
		0\\
		i_1\\
		\vdots\\
		i_6
	\end{pmatrix}.
\end{equation}
Given a constant coefficient $c$ of $f(x)$, from the above, the condition for $f(x)$ to have solutions is that $b_0(c)=0$.
If solvable, then the obtained $(i_1,i_2,i_3,i_4,i_5,i_{6})$ in \eqref{eq:21} is exactly the solution of $f(x)$ with $i_0$ either zero or one, i.e.,
$f(w_{\sum_{j=1}^6i_j\cdot 2^j})=0$ and $f(w_{\sum_{j=1}^6i_j\cdot 2^j}+1)=0$.
More examples are shown in TABLE~\ref{tab:0}, where all missing elements in matrices are zeros.



\section{Complexity Analysis}\label{sec:3}
This section analyzes the computational complexity of the above method.
The computational complexity of checking whether the reduced quadratic polynomial $f(x)$ has solutions (i.e., Theorem~\ref{thm:1}) is as follows:
\begin{lemma}\label{lem:01}
	Given the matrix pair $(P, I_0)$ for $\mathbb{F}_{2^m}$, which is shown in \eqref{eq:13}, only $m-1$ XORs are required to determine whether $f(x)\in \mathbb{F}_{2^m}[x]$ has roots in $\mathbb{F}_{2^m}$. When a parallel structure is used, it takes only  $\lceil \log_2 m\rceil$ XORs time to perform. 
	
	\begin{proof}
		According to Theorem~\ref{thm:1}, one can know which row of $I_0$ is an all-zero row.
		Then by multiplying the row of $P$ by $(b_0(c),b_1(c),...,b_{m-1}(c))^{\mathrm{T}}$, one can determine whether the condition in Theorem~\ref{thm:1} holds.
		This results in at most $m-1$ XORs.
		When using a parallel structure, only $\lceil \log_2 m \rceil$ XOR operations are required through paired merging. 
		This completes the proof.
	\end{proof}
\end{lemma}

The following lemma provides the computational complexity of solving $f(x)=0$ (i.e., Theorem~\ref{the:1}). 
\begin{lemma}
	Given the matrix pair $(P, I_0)$ for $\mathbb{F}_{2^m}$, which is shown in \eqref{eq:13}, if $f(x)\in \mathbb{F}_{2^m}[x]$ has roots in $\mathbb{F}_{2^m}$, the two distinct roots of $f(x)$ can be obtained through up to $m^2-2m+1$ XORs. When a parallel structure is used, it takes only  $\lceil \log_2 m\rceil$ XORs time to perform.
	
	\begin{proof}
		According to \eqref{eq:17}, $(i_1, i_2, ..., i_{m-1})$ is calculated by
{	\small	\begin{equation}
			\begin{aligned}
				&(P_0, P_1, ..., P_{m-1})\cdot 	\begin{pmatrix}
					b_0(c)\\
					b_1(c)\\
					\vdots\\
					b_{m-1}(c)
				\end{pmatrix}  = I_0 \cdot 
				\begin{pmatrix} 
					i_1\\
					\vdots\\
					i_{m-1}
				\end{pmatrix} \\
				\Rightarrow \quad &\sum_{j=0}^{m-1} P_j  \cdot b_j(c)  = I_0 \cdot 
				\begin{pmatrix} 
					i_1\\
					\vdots\\
					i_{m-1}
				\end{pmatrix} \;,
			\end{aligned}
		\end{equation}}
		where $(P_0, P_1, ..., P_{m-1})=P$ with each $P_j$ being the column vector of size $m\times 1$.
		Note that $I_0$ does not generate any operation, as it is the $m\times (m-1)$ matrix obtained by inserting one all-zero row to an $(m-1)\times (m-1)$ identity matrix. 
		Furthermore, the row corresponding to the all-zero row of $I_0$ does not need to participate in the calculation.
		Then the total number of XORs required is at most
		$
		(m-1)\cdot (m-1).
		$
		Through paired merging, $(i_1, i_2, ..., i_{m-1})$ can also be obtained by summing the $m\times 1$ column vectors (i.e., $P_0,P_1,...,P_{m-1}$) $\lceil \log_2 m \rceil$ times. Therefore, when a parallel structure is used, it only takes the time to perform $\lceil \log_2 m\rceil$ XORs.
		This completes the proof.
	\end{proof}
\end{lemma}

In the following, the method proposed in this paper is compared with other methods in~\cite{chien1964,chen1982formulas,cherly1998solving,walker1999new}. 
The third column of TABLE~\ref{tab:1} provides the formula expression for one root $x_0$ (clearly, the other root is $x_1=x_0+1$), and the fourth column shows the number of operations required for each method.
Note that for the case where $m$ is an even number in \cite{chen1982formulas}, only the operations for calculating $Tr_4$(c) were counted, although the remaining computational costs are substantial.
In \cite{cherly1998solving}, the field element $u$ can be found in advance, and its related results can be pre-calculated.
Furthermore, the method in \cite{walker1999new} needs to transform $f(x)$ into the form of $x^2+t_0x+t_1$, where $t_0\in \mathbb{F}_{2^m}\setminus \mathbb{F}_{2^{m/2}}$, $t_1\in \mathbb{F}_{2^m}$, and then apply the corresponding formula to solve. This generates a large number of operations that can be pre-calculated.
All pre-calculated operation quantities are not included in TABLE~\ref{tab:1}.
Due to the fact that field addition is the simplest and most efficient implementation among all operations~(a field addition over $\mathbb{F}_{2^m}$ is equivalent to $m$ XORs), one can see that the proposed method not only has the uniform form for arbitrary $m$, but also operates most efficiently, requiring only $m^2-2m+1$ XORs.

Particularly, previous methods have relied on checking whether $Tr_2(c)$ is zero to determine if $f(x)$ has solutions in $\mathbb{F}_{2^m}$, where the definition of $Tr_2(c)$ is shown in TABLE~\ref{tab:1}.
This results in $m-1$ exponentiations and $m-1$ field additions. As a comparison, the method in Theorem~\ref{thm:1} is more efficient, requiring only $m-1$ XORs, as analyzed in Lemma~\ref{lem:01}.

\section{Conclusion}\label{sec:con}
In this paper, a new formula for solving quadratic equations over binary extension fields is proposed.
Given a specific matrix pair for $\mathbb{F}_{2^m}$, the proposed formula requires at most $m^2-2m+1$ XORs to obtain the roots of the solvable quadratic equation $x^2+x+c\in \mathbb{F}_{2^m}[x]$. 
Particularly, it takes only  $\lceil \log_2 m\rceil$ XORs time to perform when a parallel structure is used.
Compared to other methods known in the literature, the proposed method has the highest computational efficiency.

%

\bibliographystyle{IEEEtran}
\bibliography{ref}

\end{document}